\documentclass[runningheads]{llncs}

\usepackage[parfill]{parskip}    
\usepackage[T1]{fontenc}
\usepackage{amsfonts}
\usepackage{amsmath}
\usepackage{float}
\usepackage{graphicx}
\makeatletter
\def\subsubsection{\@startsection{subsubsection}{3}%
  \z@{.5\linespacing\@plus.7\linespacing}{.1\linespacing}%
  {\normalfont\itshape}}
\makeatother

\begin{document}
\title{Shape-guided Conditional Latent Diffusion Models for Synthesising Brain Vasculature}
\titlerunning{Latent Diffusion Models for Synthesising Brain Vasculature}
%
\author{Yash Deo\inst{1} \and
Haoran Dou\inst{1} \and
Nishant Ravikumar\inst{1,2} \and 
Alejandro F.~Frangi\inst{1,2,3,4,5} \and 
Toni Lassila\inst{1,2}
}
\authorrunning{Y. Deo et al.}
%
\institute{Centre for Computational Imaging and Simulation Technologies in Biomedicine (CISTIB), School of Computing and School of Medicine, University of Leeds, Leeds, UK 
\and
NIHR Leeds Biomedical Research Centre (BRC), Leeds, UK
\and
Alan Turing Institute, London, UK
\and
Medical Imaging Research Center (MIRC), Electrical Engineering and Cardiovascular Sciences Departments, KU Leuven, Leuven, Belgium
\and
Division of Informatics, Imaging and Data Science, Schools of Computer Science
and Health Sciences, University of Manchester, Manchester, UK
}
\maketitle              
\begin{abstract}
The Circle of Willis (CoW) is the part of cerebral vasculature responsible for delivering blood to the brain. Understanding the diverse anatomical variations and configurations of the CoW is paramount to advance research on cerebrovascular diseases and refine clinical interventions. However, comprehensive investigation of less prevalent CoW variations remains challenging because of the dominance of a few commonly occurring configurations. We propose a novel generative approach utilising a conditional latent diffusion model with shape and anatomical guidance to generate realistic 3D CoW segmentations, including different phenotypical variations. Our conditional latent diffusion model incorporates shape guidance to better preserve vessel continuity and demonstrates superior performance when compared to alternative generative models, including conditional variants of 3D GAN and 3D VAE. We observed that our model generated CoW variants that are more realistic and demonstrate higher visual fidelity than competing approaches with an FID score 53\% better than the best-performing GAN-based model. 

\keywords{Image Synthesis  \and Deep Learning \and Brain Vasculature \and Vessel Synthesis \and Diffusion \and Latent Diffusion}
\end{abstract}
\section{Introduction}

The Circle of Willis (CoW) comprises a complex network of cerebral arteries that plays a critical role in the supply of blood to the brain. The constituent arteries and their branches provide a redundant route for blood flow in the event of occlusion or stenosis of the major vessels, ensuring continuous cerebral perfusion and mitigating the risk of ischaemic events~\cite{Lin2022}. However, the structure of the CoW is not consistent between individuals and dozens of anatomical variants exist in the general population ~\cite{C2,C1}. Understanding the differences between these variants is essential to study cerebrovascular diseases, predict disease progression, and improve clinical interventions. Previous studies have attempted to classify and describe the anatomical variations of CoW using categorisations such as the Lippert and Pabst system~\cite{C2,C1}. However, more than 80\% of the general population has one of the three most common CoW configurations~\cite{C3}. 
The study of anatomical heterogeneity in CoW is limited by the size of available angiographic research data sets, which may only contain a handful of examples of all but the most common phenotypes. The goal of this study is to develop a generative model for CoW segmentations conditioned on anatomical phenotype. Such a model could be used to generate large anatomically realistic virtual cohorts of brain vasculature, and the less common CoW phenotypes can be augmented and explored in greater numbers. Synthesised virtual cohorts of brain vasculature may subsequently be used for training deep learning algorithms on related tasks (e.g. segmenting brain vasculature, classification of CoW phenotype, etc.), or performing in-silico trials.

Generative adversarial networks (GANs)~\cite{GAN} and other generative models have demonstrated success in the synthesis of medical images, including the synthesis of blood vessels and other anatomical structures. However, to the best of our knowledge, no previous study has explored these generative models for synthesising different CoW configurations. Additionally, no previous study has explored the controllable synthesis of different CoW configurations conditioned on desired phenotypes. The synthesis of narrow tubular structures such as blood vessels using conventional generative models is a challenge. Our study builds upon the foundations of generative models in medical imaging and focusses on utilising a conditional latent diffusion model to generate visually realistic CoW configurations with controlled anatomical variations (i.e., by conditioning relevant anatomical information such as CoW phenotypes). Medical images like brain magnetic resonance angiograms (MRA's) tend to be high-dimensional and as a result are prohibitively memory intensive for generative models. Diffusion models and latent diffusion models (LDM) have recently been used for medical image generation~\cite{D0} and have been shown to outperform GANs in medical image synthesis~\cite{D1}. Diffusion models have also been successfully used to generate synthetic MRIs~\cite{D4,D2,D3} but to the best of our knowledge there are no studies that use latent diffusion models are diffusion models to generate synthetic brain vasculature.

We propose a conditional latent diffusion model that learns latent embeddings of brain vasculature and, during inference, samples from the learnt latent space to synthesise realistic brain vasculature. We incorporate class, shape, and anatomical guidance as conditioning factors in our latent diffusion model, allowing the vessels to retain their shape and allowing precise control over the generated CoW variations. The diffusion model is conditioned to generate different anatomical variants of the posterior cerebral circulation. We evaluate the performance of our model using quantitative metrics such as multiscale structural similarity index (MS-SSIM) and Fr'echet inception distance (FID). Comparative analyses are conducted against alternative generative architectures, including a 3D GAN and a 3D variational auto-encoder (VAE), to assess the superiority of our proposed method in reproducing CoW variations.
 
\section{Methodology}
 

\textbf{Data and Pre-processing.} We trained our model on the publicly available IXI dataset ~\cite{IXI} using the 181 3T MRA scans acquired at the Hammersmith Hospital, London. Images were centred, cropped from $512\times512\times100$ to $256\times256\times100$, and the intensity normalised. We then used a Residual U-net~\cite{Kerfoot18} to extract vessel segmentations from the MRA. The authors manually labelled each case with the presence / absence of one or both peripheral communicating arteries in the CoW. Class 1 includes cases where both the peripheral communication arteries are present (PComA), Class 2 includes cases with only one PComA, while Class 3 includes cases where both PComAs are absent.

\textbf{Latent Diffusion Model.} Recent advances in diffusion models for medical image generation have achieved remarkable success. Diffusion models define a Markov chain of diffusion steps to add random Gaussian noise to the observed data sequentially and then learn to reverse the diffusion process to construct new samples from the noise. Although effective, vanilla diffusion models can be computationally expensive when the input data is of high dimensionality in image space ($256\times256\times100$ in our study). Hence, we employ the latent diffusion model (LDM), comprising a pretrained autoencoder and a diffusion model. The autoencoder learns a lower-dimensional latent embedding of the brain vasculature, while the diffusion model focusses on modelling the high-level semantic representations in the latent space efficiently. 


Following~\cite{D1}, the diffusion process can be defined as forward and reverse Markov chains, where the forward process iteratively transforms the data $x_0$ (i.e. the latent features from the autoencoder in our approach) into a standard Gaussian $X_T$ as following:
\begin{equation*}
    q \left( \mathbf{x}_{1:T} | \mathbf{x}_{0} \right) = \prod_{t=1}^{T} q \left( \mathbf{x}_t | \mathbf{x}_{t-1} \right), q \left( \mathbf{x}_{t} | \mathbf{x}_{t-1} \right) := \mathcal{N} \left( \mathbf{x}_t ; \sqrt{1-\beta_t} \mathbf{x}_{t-1}, \beta_{t}\mathbf{I} \right)
\end{equation*}
where $q \left( \mathbf{x}_{t} | \mathbf{x}_{t-1} \right)$ is the transition probability at the time step $t$ based on the noise schedule $\beta_{t}$. Therefore, the noisy data $\mathbf{x}_t$ can be formulated as $q \left( \mathbf{x}_{t} | \mathbf{x}_{0} \right) = \mathcal{N} \left( \mathbf{x}_t ; \sqrt{\bar{\alpha}_t} \mathbf{x}_{0}, (1-\bar{\alpha}_t)\mathbf{I} \right)$, where $\alpha_t := 1-\beta_t, \bar{\alpha}_t := \prod_{s=1}^t \alpha_s$. 

The reverse process, achieved via a deep neural network parameterised by $\theta$, can then be defined as:
\begin{equation*}
    p_\theta \left( \mathbf{x}_{0} | \mathbf{x}_{T} \right) = p \left( \mathbf{x}_T \right) \prod_{t=1}^{T} p_\theta \left( \mathbf{x}_{t-1} | \mathbf{x}_{t} \right), p_\theta \left( \mathbf{x}_{t-1} | \mathbf{x}_{t} \right) := \mathcal{N} \left( \mathbf{x}_{t-1} ; \mathbf{\mu}_\theta \left( \mathbf{x}_{t}, t\right), \mathbf{\Sigma}_\theta \left( \mathbf{x}_t, t \right) \right)
\end{equation*}
The simplified evidence lower bound (ELBO) loss to optimise the diffusion model by Ho~\textit{et al.}~\cite{D1} can be formulated as a score-matching task where the neural network predicts the actual noise $\epsilon$ added to the observed data. The resulting loss function is
$
\mathcal{L}_{\theta} := \mathbb{E}_{\textbf{x}_0, t, C, \epsilon \sim \mathcal{N} \left( 0,1 \right)} \left[ \left\| \epsilon - \epsilon_\theta \left(x_t, t, C\right)\right\|^2\right]
$
where $C$ is the condition in conditional generation.

We pretrained a multitask attention-based autoencoder using a combination of L1 loss and Dice loss. The encoder transforms the brain image $K_0$ into a compact latent representation $x_0$ with dimensions of $256\times256\times1$. Once the compression model is trained, the latent representations from the training set serve as inputs to the diffusion model for further analysis and generation.

We employ a model with a U-net-based architecture as the diffusion model. Our model has 5 encoding blocks and 5 decoding blocks with skip connections between the corresponding encoding and decoding blocks. We replace the simple convolution layers in the encoding and decoding blocks with a residual block followed by a multihead attention layer to limit information loss in the latent space. Each encoding and decoding block takes the class category (based on CoW phenotypes) as an additional conditional input, while, only the decoding blocks take shape and anatomy features as additional conditional inputs.

\begin{figure}[t!]
\centering
\includegraphics[width=0.85\textwidth]{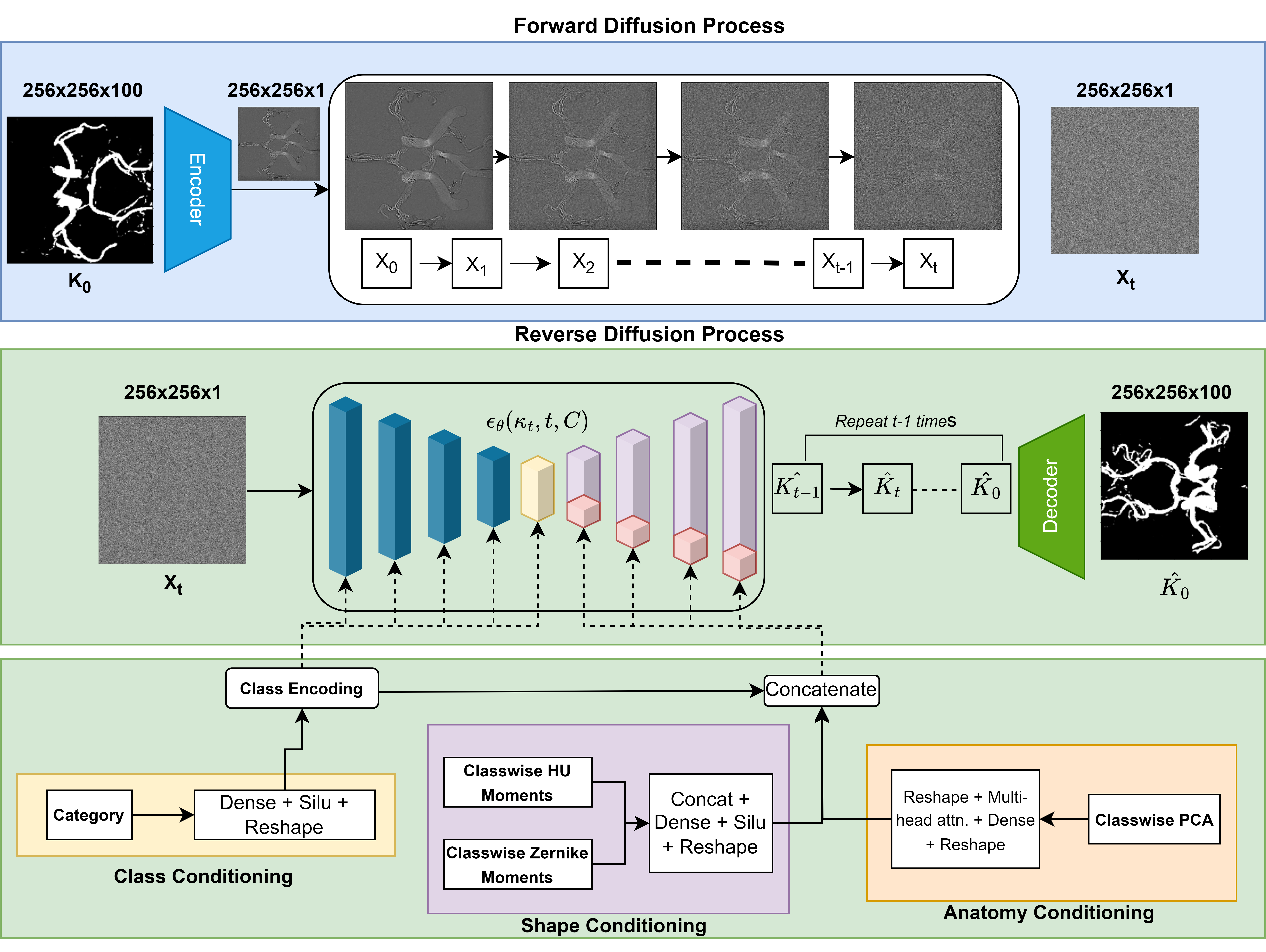}
\caption{Overview of the latent diffusion process.} 
\label{fig:1}
\end{figure}


\textbf{Shape and Anatomy Guidance.} Angiographic medical images exhibit intricate anatomical structures, particularly the small vessels in the peripheral cerebral vasculature. Preserving anatomical integrity becomes crucial in the generation of realistic and accurately depicted vessels. However, diffusion models often face challenges in faithfully representing the anatomical structure, which can be attributed to their learning and sampling processes that are heavily based on probability density functions ~\cite{DM12}. Previous studies have demonstrated that the inclusion of geometric and shape priors can improve performance in medical image synthesis~\cite{SP1,SP2}. Additionally, latent space models are susceptible to noise and information loss within the latent space. To this end, we incorporate shape and anatomy guidance to improve the performance of our CoW generation.

The shape guidance component involves incorporating class-wise Hu and Zernike moments as conditions during model training~\cite{H1,Z1}. This choice stems from the nature of our image dataset, which comprises both vessel and background regions. By including these shape-related moments as conditions, we aim to better preserve vascular structures within the synthesised images. Hu and Zernike moments are a set of seven invariant moments and a set of orthogonal moments, respectively, commonly used for shape analysis. These moments are typically computed on greyscale or binary images. To incorporate the Hu and Zernike moments as conditions, we first calculate and concatenate these moments for each class. An embedding layer comprising a dense layer with a SiLU activation function~\cite{Silu} and a reshape layer is then introduced to ensure that the data are reshaped into a suitable format for integration as a condition within the decoding branches.

\begin{figure}[h!]
\centering
\includegraphics[width=0.9\textwidth]{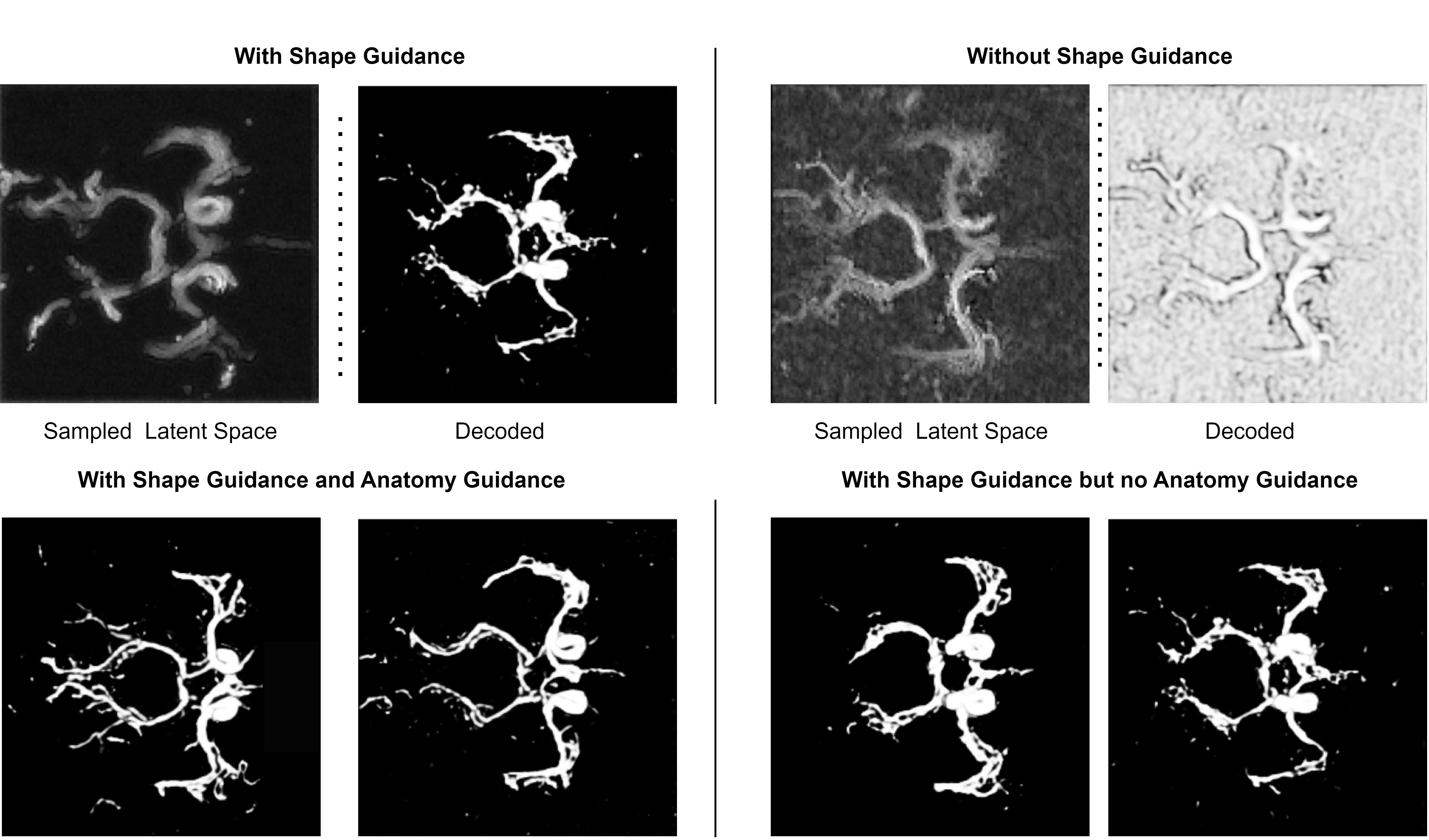}
\caption{Row 1: Comparison of output of the latent diffusion network with and without using shape guidance as conditional input. In each column, the image on the left shows the output of our latent diffusion model and the image on the right shows the result of passing the output through the pretrained decoder and obtaining the Maximum Intensity Projection (MIP); Row 2: compares the output of the network with and without using anatomy guidance as conditional input. The generated images displayed on the right, which are produced without the incorporation of anatomy guidance, consistently exhibit a similar variation of the circle of Willis. Conversely, the images presented on the left, which are generated with the inclusion of anatomy guidance, demonstrate a greater degree of realism and variability in the synthesised circle of Willis variations. } \label{fig:2}
\end{figure}

To further enhance the performance of our model, we incorporate anatomy guidance using principal component analysis (PCA) on images from each class. As the majority branches within the CoW exhibit a consistent configuration with minor variations attributed to the presence or absence of specific branches, the model tends to capture an average or mean representation of the CoW and generates synthetic images with very little variation between them. This characteristic becomes significant due to the limited number of images available per class. To address this, we use PCA components as conditions to enable the model to discern distinctive features specific to each class. We extract seven principal components along with the mean component for each class, concatenate them, and reshape the data. The resulting features are then passed through a multi-head attention block, followed by a dense layer and another reshape operation for integration into the decoding branches.

Fig.~\ref{fig:2} shows the effect of incorporating shape moments and PCA as conditions in our diffusion process. By incorporating shape and anatomy guidance conditions during the training of our diffusion model, we leverage specific features and knowledge related to the vessel structures and the general anatomy of the images. This approach promotes the generation of more realistic images, contributing to an improved anatomical fidelity.

\section{Experiments and results}


\textbf{Implementation Details.} All models were implemented in TensorFlow 2.8 and Python 3. For the forward diffusion process we use a linear noise schedule with 1000 time steps. The model was trained for 2000 epochs with a learning rate of 0.0005 on a Nvidia Tesla T4 GPU and 38 Gb of RAM with Adam optimiser.

\begin{figure}[h!]

\includegraphics[width=\textwidth]{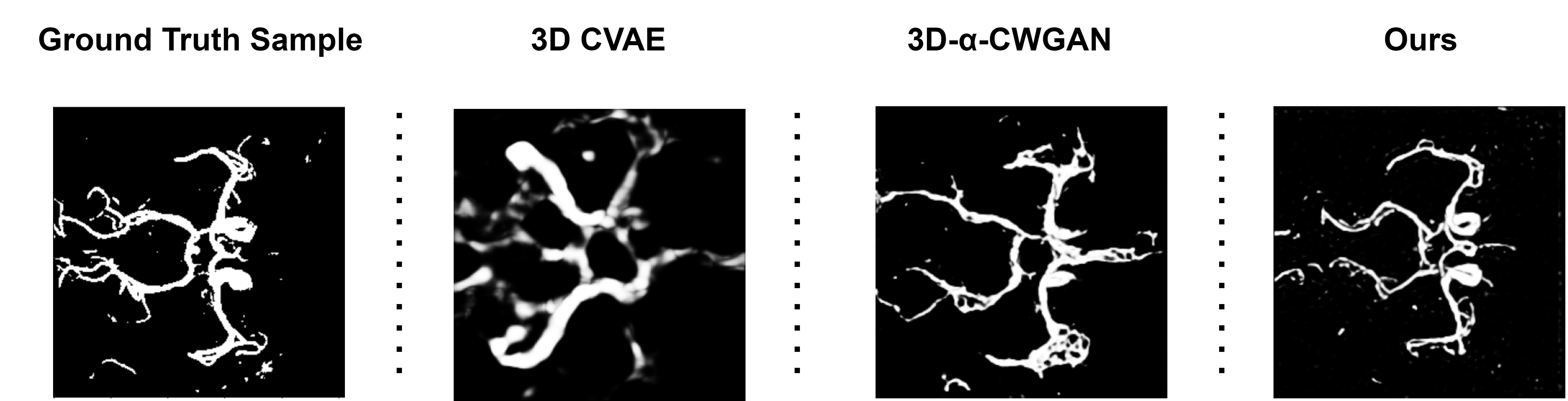}
\caption{Comparison between the maximum intensity projections (MIPs) of a real Circle of Willis(CoW) against those synthesised with 3D CVAE, 3D-$\alpha$-WGAN, and our model.   } \label{fig:4}
\end{figure}

\textbf{Results and Discussion.} To assess the performance of our model, we compared it against two established conditional generative models: 3D C-VAE~\cite{VAE} and a 3D-$\alpha$-WGAN~\cite{AG} along with a vanilla LDM and an LDM with shape guidance. We use the FID score to measure the realism of the generated vasculature. To calculate FID we used a pre-trained InceptionV3 as a feature extractor. A lower FID score indicates higher perceptual image quality. In addition, we used MS-SSIM and 4-G-R SSIM to measure the quality of the generated images~\cite{Cl1,Cl2}. MS-SSIM and 4-G-R SSIM are commonly used to assess the quality of synthesised images. Typically, a higher score is indicative of better image quality, implying a closer resemblance between the synthesised CoW and the ground truth reference. MS-SSIM and 4-G-R SSIM were calculated over 60 synthesised CoW cases for each model. Table 1 presents the evaluation scores achieved by our model, 3D CVAE, and the 3D-$\alpha$-WGAN and the above metrics. As seen in Table \ref{table:1}, our model demonstrates a better FID score, suggesting that the distribution of CoW variants synthesised by our model is closer to that observed in real CoW data, compared to the other models. Additionally, our model achieves higher MS-SSIM and 4-G-R SSIM scores compared to the other methods. These higher scores indicate better image quality, implying that the generated CoW samples resemble the real CoW images more closely.  Fig.~\ref{fig:4} provides a qualitative comparison among the generated samples obtained from the three models to provide additional context to the quantitative results presented in Table 1. As the output of each model is a 3D vascular structure, maximum intensity projections (MIP) over the Z-axis which condense the volumetric representation into a 2D plane are used to visually compare the synthesised images.

\begin{table}[ht!]
\centering
\caption{Quantitative evaluation of Synthetic CoW vasculature}
\label{table:1}
\begin{tabular}{l|c|c|c}
Model        &  FID $\downarrow$ & MS-SSIM $\uparrow$ & 4-G-R SSIM $\uparrow$ \\ 
\hline
3D CVAE              & $52.78$  & $0.411$ & $0.24$  \\
3D-$\alpha$-WGAN   & $12.11$  & $0.53$ & $0.41$ \\
LDM   & $176.41$  & $0.22$ & $0.13$ \\
LDM + Shape Guidance   & $8.86$  & $0.58$ & $0.47$ \\
Ours (LDM + Shape \& Anatomy Guidance)               &  $5.644$ & $0.61$  & $0.51$ \\ 
\end{tabular}
\end{table}

Fig.~\ref{fig:4} reveals that the 3D CVAE model can only generate a limited number of major vessels with limited details. On the other hand, although the 3D-$\alpha$-WGAN model produces the overall structure of the CoW, it exhibits significant anatomical discrepancies with the presence of numerous phantom vessels. On the contrary, our model demonstrates a faithful synthesis of the majority of CoW, with most vessels identifiable. To generate variations of the CoW based on the presence or absence of the posterior communicating artery, our latent diffusion model uses class-conditional inputs where the classes represent different CoW phenotypes. Consequently, to demonstrate the class-conditional fidelity of the proposed approach, we also evaluate the model's performance in a class-wise manner. The qualitative performance of our model for different classes, compared to real images belonging to those classes, is shown in Fig.~\ref{fig:4}

\begin{figure}[h!]
\centering
\includegraphics[width=0.85\textwidth]{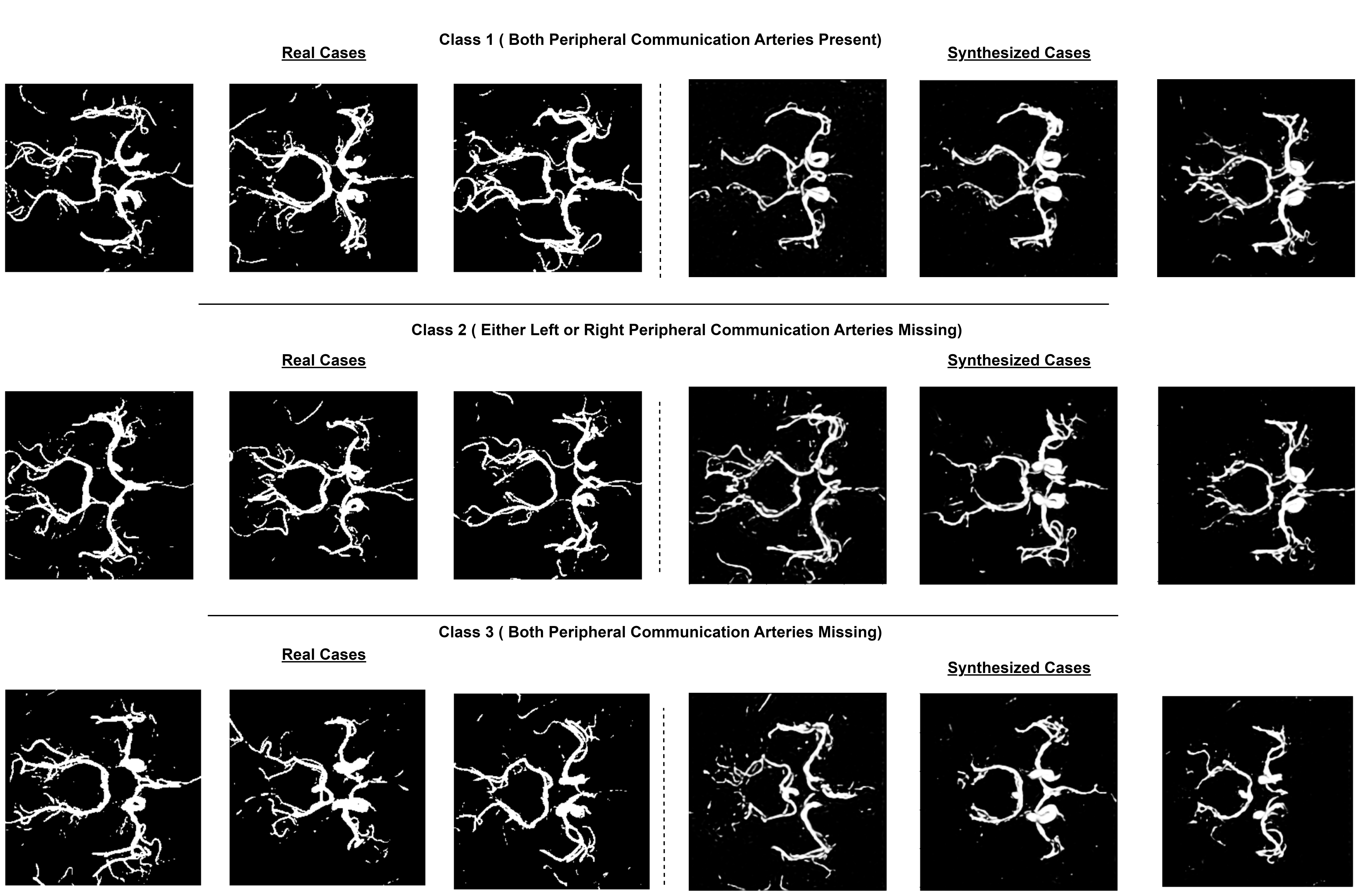}
\caption{Comparison between the real and synthesised maximum intensity projections (MIPs) for each of the three classes   } \label{fig:5}
\end{figure}

\begin{table}[ht!]
\centering
\caption{Quantitative class-wise evaluation of Generated CoW vasculature}
\label{table:2}
\begin{tabular}{l|c|c|c}
Class     &  FID Score $\downarrow$ & MS-SSIM $\uparrow$ & 4-G-R SSIM $\uparrow$ \\ 
\hline
Class 1   & $4.41$  & $0.65$ & $0.65$\\
Class 2   & $3.88$  & $0.52$ & $0.52$\\
Class 3   & $7.63$ & $0.41$ & $0.41$\\ 
Overall   & $5.64$ & $0.61$  & $0.51$\\ 
\end{tabular}
\end{table} 

The results presented in Fig.~\ref{fig:5} demonstrate the performance of our model in generating realistic variations of the Circle of Willis. Particularly notable is the model's proficiency in producing accurate representations for classes 1 and 2, surpassing its performance in class 3 due to the limited sample size of the latter. Our model excels in synthesising the posterior circulation and the middle cerebral arteries, showing remarkable fidelity to anatomical structures. However, it faces challenges in effectively generating continuous representations of the anterior circulation. Further investigation and refinement may be required to enhance the model's ability in this specific aspect. In addition to the visual assessment, we also compute class-wise FID scores, along with the MS-SSIM and 4-G-R SSIM scores. These quantitative evaluations serve to provide a more comprehensive understanding of the model performance with respect to each class. The class-wise performance scores shown in Table~\ref{table:2} are consistent with our observations from Fig.~\ref{fig:5}, that the model's performance for class 3 is worse than its performance on classes 1 and 2.



\section{Conclusion}
We proposed a latent diffusion model that used shape and anatomy guidance to generate realistic CoW configurations. Quantitative qualitative results showed that our model outperformed existing generative models based on a conditional 3D GAN and a 3D VAE. Future work will look to enhance the model to capture wider anatomical variability and improve synthetic image quality. 

\section*{Acknowledgement}

This research was partially supported by the National Institute for Health and Care Research (NIHR) Leeds Biomedical Research Centre (BRC) and the Royal Academy of Engineering Chair in Emerging Technologies (CiET1919/19).

%

\bibliographystyle{splncs04}
\bibliography{mybib}
\end{document}